\documentclass[preprint,superscriptaddress,amsmath,amssymb,aps,notitlepage,floatfix,longbibliography]{revtex4-2}
\usepackage{graphicx}
\usepackage{comment}
\usepackage{caption}
\usepackage{upgreek}
\usepackage{gensymb}
\usepackage{float}
\usepackage{bm}
\usepackage{xcolor}
\usepackage{physics}
\usepackage{multirow}
\usepackage{array}
\usepackage{makecell}
\usepackage[normalem]{ulem}

\begin{document}

\title{Photo-thermal origin of pulse laser induced orientation of crystallographic $c$ axis in Tellurium thin films}

\author{Arata Mitsuzuka}
\affiliation{Department of Physics, The University of Tokyo, Tokyo 113-0033, Japan}

\author{Yuta Kobayashi}
\email[]{yuta.kobayashi@phys.s.u-tokyo.ac.jp}
\affiliation{Department of Physics, The University of Tokyo, Tokyo 113-0033, Japan}

\author{Takuto Hiraoka}
\affiliation{Department of Physics, The University of Tokyo, Tokyo 113-0033, Japan}

\author{Masashi Kawaguchi}
\affiliation{Department of Physics, The University of Tokyo, Tokyo 113-0033, Japan}

\author{Masamitsu Hayashi}
\email[]{hayashi@phys.s.u-tokyo.ac.jp}
\affiliation{Department of Physics, The University of Tokyo, Tokyo 113-0033, Japan}
\affiliation{Trans-Scale Quantum Science Institute, The University of Tokyo, Bunkyo, Tokyo 113-0033, Japan}

\date{\today}

\begin{abstract}
Recent studies have shown that the orientation of crystallographic $c$ axis of Tellurium thin films can be controlled using picosecond long laser pulses.
This method provides spatially programmable control of the crystal orientation and is therefore highly attractive for practical applications in functional optical and electronic devices.  
Previously, it was suggested that laser-induced selective melting and recrystallization can cause the laser-induced reorientation.
However, this interpretation remains inconclusive due to limited data.
To clarify the mechanism, here we systematically study Te samples under different irradiation conditions.
We find that the threshold fluence for inducing optical reorientation depends on the number of laser pulses.
The results agrees well with a minimal kinetic model based on the Arrhenius law.
Using the model developed, we investigate the condition required to control the optic axis in other two-dimensional materials, such as black phosphorus, WTe$_2$, and SnSe. 
These findings provide a guide for developing functional electro-optical devices based on anisotropic materials.
\end{abstract}
\maketitle

\clearpage
\section{Introduction}
Tellurium is an elemental semiconductor with a chain-like crystal structure that gives rise to strong uniaxial anisotropy.\cite{adenis1989actacryst,rikken2019prb}. 
Key properties of Te, such as its refractive index\cite{zhang2023nanomater} and carrier mobility\cite{tong2020ncomm}, strongly depend on the orientation of the crystallographic $c$ axis.
Controlling the orientation of the $c$ axis is therefore important for a wide range of applications, including polarization-sensitive optical devices\cite{yu2024anisotropic}, transistors\cite{tang2025wafer}, and mid-infrared light-emitting diodes\cite{zhang2025dynamically}.

Recently, it was reported that the $c$ axis of Te can be reoriented by linearly polarized laser pulses. 
Upon irradiation, the $c$ axis aligns perpendicular to the polarization of the incident laser pulses, enabling on-demand control and spatially resolved patterning of its orientation \cite{kobayashi2026nanolett}.
Despite the potential impact of this technique, the microscopic mechanism of the reorientation process remains unclear.
Previously, a selective melting and recrystallization process driven by anisotropic optical absorption was proposed as a possible origin.
However, no physical model has yet been established that can quantitatively account for the observed behavior.
In particular, the relationship between the mechanism governing orientation control and the intrinsic properties of Te remains elusive; furthermore, it is uncertain whether the experimentally observed reorientation is driven solely by thermal effects.

\begin{figure}[ht]
    \begin{center}
    \includegraphics[width=0.8\linewidth]{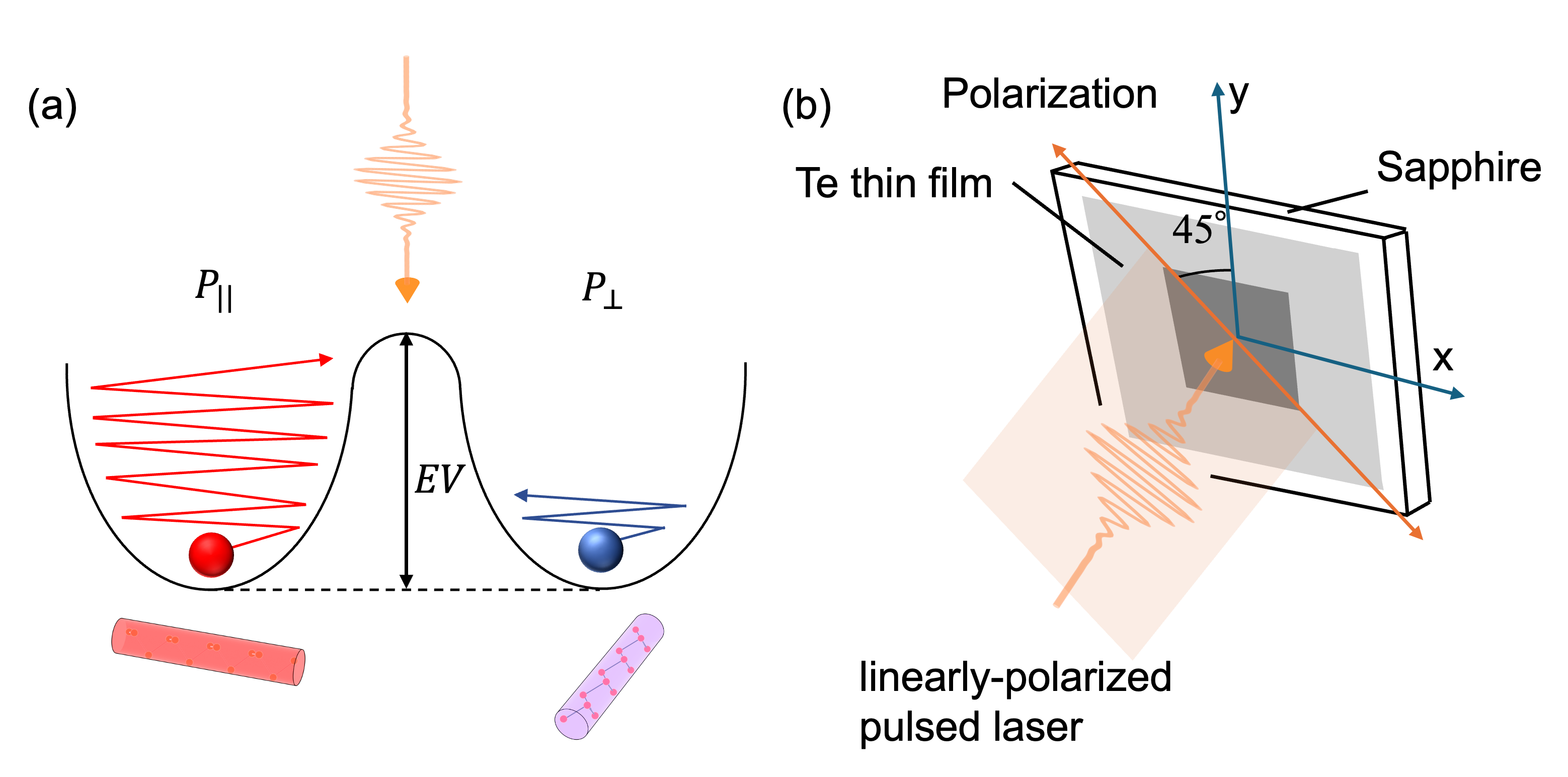}
    \caption{(a) Schematic of the proposed minimal kinetic model. A linearly polarized laser pulse induces orientation-dependent heating through anisotropic optical absorption. The resulting state-dependent effective temperature modifies the probability of overcoming the activation barrier $EV$ leading to selective reorientation.
            (b) Illustration of the pulse laser irradiation. The tellurium thin film on the sapphire substrate was irradiated at normal incidence with linearly polarized laser pulses. The laser polarization was oriented 45$^\circ$ from the $y$-axis. Counterclockwise direction is taken as positive when viewed from the \(+z\) side.
        }
    \label{fig:schematic}
    \end{center}
\end{figure}

Here, we systematically studied pulse-laser-induced $c$ axis reorientation under various number of pulses.
The refractive index of the irradiated area was measured to determine the degree of $c$ axis alignement. 
We find that the threshold to induce reorientation depends on the number of pulses in an exponential way.
This feature is well reproduced by a minimal kinetic model based on the Arrhenius law, indicating that the $c$ axis control can be understood predominantly in terms of thermal effects [see Fig.~\ref{fig:schematic}(a)].
That is, due to linear dichroism of Te, the laser pulse selectively melts Te grains with $c$ axis parallel to laser polarization, upon which random reorientation takes place.
After application of a large number of pulses, grains with $c$ axis perpendicular to the laser pulse remain, thus allowing alignment of the average orientation.
As the effect is induced by anisotropic absorption of pulse energy, it is readily applicable to a wide range of materials. 
We show that black phosphorus, WTe$_2$, and SnSe are also promising candidates for laser-induced control of the optic axis.

The outline of this paper is as follows.
First, we describe sample preparation and the experimental setup.
We then present the reflectance and transmittance measurement results performed to determine the refractive index of the Te films irradiated with pulse laser under different conditions.
To extract the refractive index, we take into account both multiple reflections and surface roughness of the films.
Next, we develop a minimal kinetic model for the laser induced writing of the $c$ axis and compare the calculations with the experimental results.
Finally, we extend the model to discuss the condition required to control the crystal orientation using laser pulses in other layered materials.

\section{Experimental results}

\subsection{Sample fabrication}
Films made of Te (40 nm)/Al$_2$O$_3$ (1 nm) were deposited on $c$-Al$_2$O$_3$ substrates at room temperature using molecular beam epitaxy (MBE).
The 1 nm-thick Al$_2$O$_3$ serves as a capping layer.
The Te films were irradiated with a train of linearly polarized picosecond laser pulses from the film normal.
See Fig.~\ref{fig:schematic}(b) for schematic illustration of the setup.
The center wavelength, the duration and the repetition rate of the laser were 1030 nm, 4-10 ps, and 36.6 MHz, respectively.
The laser spot was focused to a beam radius of approximately 2 $\upmu$m.
We scanned the pulse laser over an area of $100 \times 80$ $\upmu$m$^2$ of the Te film.
The scanning procedure is as follows.
The focused beam was first moved horizontally across a distance of $\sim$100 $\upmu$m.
The beam was then vertically shifted by $\sim$2 $\upmu$m.
Subsequently, the beam was moved horizontally in the opposite direction by $\sim$100 $\upmu$m, and then shifted vertically by the same distance ($\sim$2 $\upmu$m).
The scan rate was fixed during the whole scan.
We vary the scan rate (100, 200, 500, 1000, 2000 and 5000 $\upmu$m s$^{-1}$) to obtain areas with different exposure.
The corresponding number of pulses $N$ the film receives is $N$ = 1.46 $\times$ 10$^6$, 7.32 $\times$ 10$^5$, 2.93 $\times$ 10$^5$, 1.46 $\times$ 10$^5$, 7.32 $\times$ 10$^4$, 2.93 $\times$ 10$^4$ and 1.46 $\times$ 10$^4$.
The laser fluence $U$ was also changed for each scan: $U$ = 1.74, 3.48, 5.29, 6.96, 8.70, 10.4, 12.2, 13.9 and 15.7 mJ/cm$^2$.
Previous study showed that upon pulse laser irradiation, the Te $c$ axis orients perpendicular to the laser polarization\cite{kobayashi2026nanolett}.
To simplify the analyses (described later), we set the polarization direction of the pulse laser to 45 deg from the vertical ($y$-)axis: see Fig.~\ref{fig:setup} for the definition of the coordinate axis. 

\subsection{Experimental setup}
\begin{figure}[ht]
    \begin{center}
    \includegraphics[width=0.8\linewidth]{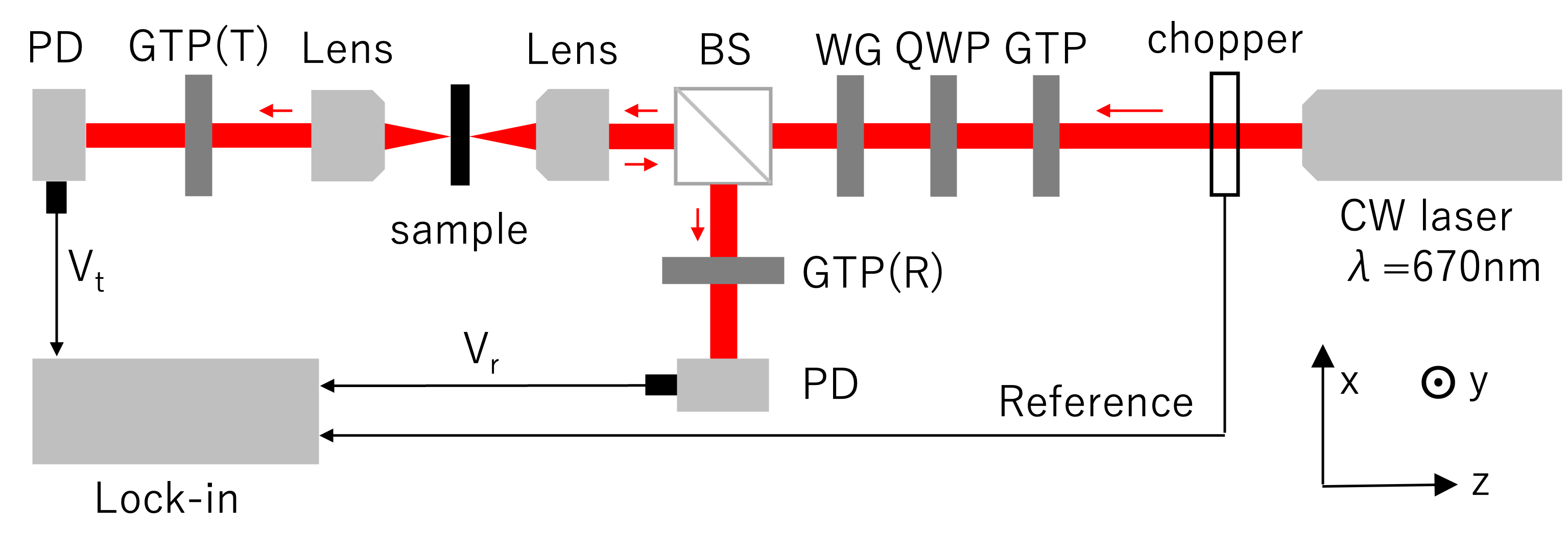}
    \caption{Illustration of the coordinate axis and the optical setup used to measure the refractive index of Te. GTP: Glan-Thompson polarizer, QWP: quarter wave plate, WG: wire-grid polarizer, BS: beam splitter, PD: photo detector. %The sample surface is defined as the \(xy\) plane, and the \(+z\) axis is taken along the surface normal from the substrate side to the air side. 
    The angle \(\theta_\mathrm{WG}\) is measured in the \(xy\) plane from the \(y\) axis, with counterclockwise direction taken as positive when viewed from the \(+z\) side.
    }
    \label{fig:setup}
    \end{center}
\end{figure}

To quantitatively determine the degree of $c$ axis alignment, defined as the fraction of the grains whose $c$ axis are reoriented by the pulse laser among all laser-irradiated grains, we exploit the optical anisotropy of Te and perform optical measurements.
Schematic configuration of the setup is shown in Fig.~\ref{fig:setup}.
A continuous wave (CW) laser (wavelength: 670 nm) was used as a light source.
The laser goes through an optical chopper, a Glan-Thompson polarizer (GTP), a quarter wave plate (QWP) and a wire-grid polarizer (WG).
The fast axis of the GTP and QWP were fixed to 0$^\circ$ and 45$^\circ$, respectively.
The angles of the fast axis of the optical comonents are defined with respect to the vertical direction: $0^\circ$ corresponds to the fast axis oriented along the $y$-axis in Fig.~\ref{fig:setup}.
The GTP linearly polarizes the laser beam and the QWP and WG set the polarization direction.
That is, the fast axis of the WG, denoted as $\theta_\mathrm{WG}$ and varied from 0$^\circ$ to 360$^\circ$, defines the polarization angle.
A non-polarizing beam splitter (BS) was placed after the WG. 
The light that transmits the BS goes through an objective lens (20x) and is focused on an area of film that was irradiated with the pulse laser.
The laser incidence was parallel to the substrate surface normal.
Note that the birefringence of the sapphire substrate does not appear under normal incidence.
The transmitted and reflected light from the film goes through a GTP, which splits 
the orthogonal polarization into orthogonal components, and then enters a photodetector (PD).
We refer to the GTP for which the transmitted and reflected light pass through as GTP(T) and GTP(R), respectively.
The output of the PD was connected to a lock-in amplifier, which was synchronized with the optical chopper (373 Hz).
The fast axis of the GTP(T) and GTP(R), defined as $\alpha$, is set to 0$^\circ$ or 90$^\circ$.

The signal from the lock-in amplifier was collected to assess the reflectance and transmittance of the film with and without pulse laser irradiation.
We also measured the reflectance and transmittance of an area of the substrate where no Te was deposited to obtain a reference signal.
To exclude artificial contributions from the optical setup, the reflectance and transmittance without any sample (air only) were measured as well.
We denote the lock-in voltage when the Te film, substrate, and air were placed in the sample stage as $V_{\mathrm{Te},\alpha}^{r/t}$, $V_{\mathrm{s},\alpha}^{r/t}$ and $V_{\mathrm{air},\alpha}^{r/t}$, respectively.
The superscript indicates the measurement geometry: $r$ for reflection and $t$ for transmission. 
The subscript denotes the analyzer angle: we label $v$ for $\alpha = 0^\circ$ and $h$ for $\alpha = 90^\circ$.
\begin{figure}[ht]
    \begin{center}
    \includegraphics[width=0.9\linewidth]{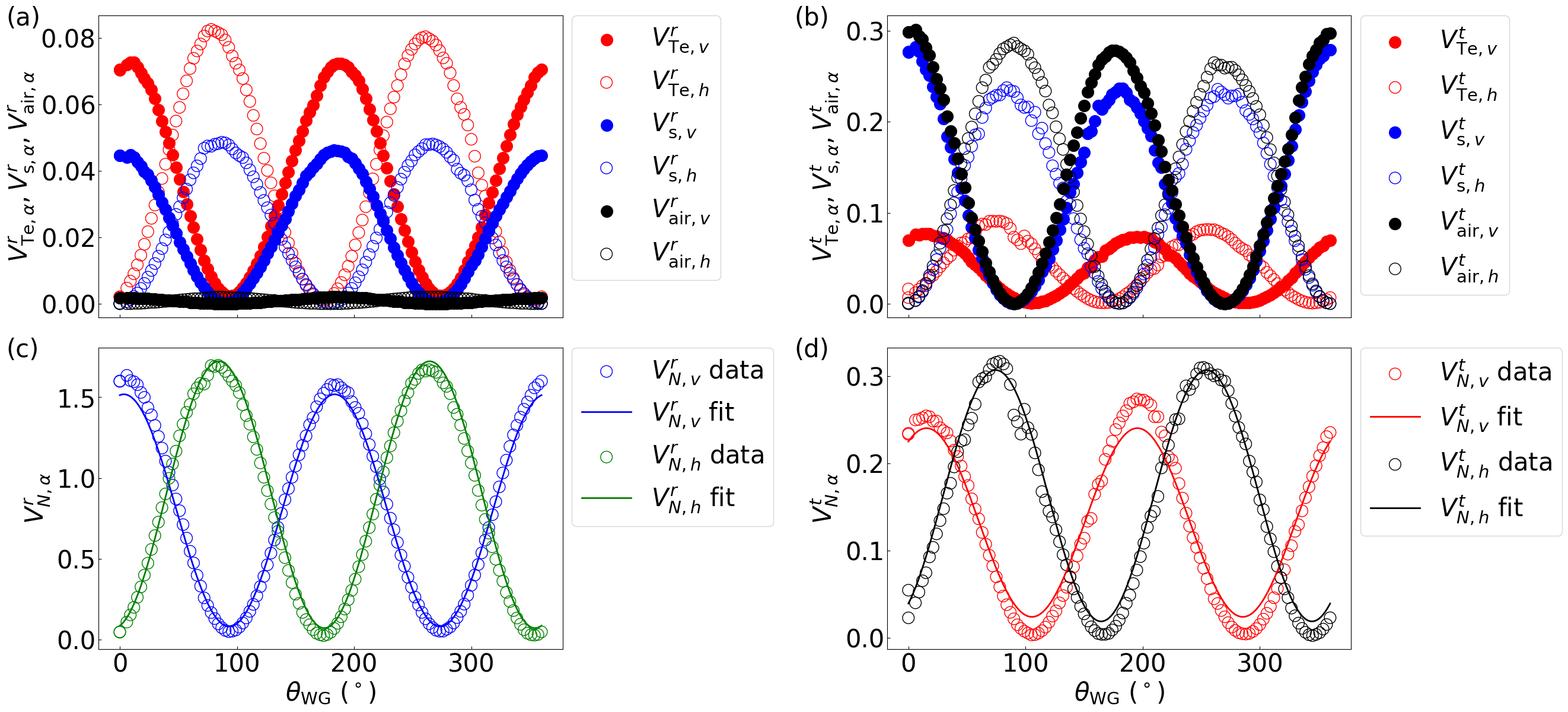}
    \caption{(a,b) WG angle $\theta_\mathrm{WG}$ dependence of the lock-in voltage obtained in the reflection (a) and transmission (b) geometries when the Te film $V_{\mathrm{Te},\alpha}^{r/t}$, the sapphire substrate $V_{\mathrm{s},\alpha}^{r/t}$ and air $V_{\mathrm{air},\alpha}^{r/t}$ are measured. (c,d) $V_{N,\alpha}^{r}$ (c) and $V_{N,\alpha}^{t}$ (d) with $\alpha = h, v$ are plotted against $\theta_\mathrm{WG}$ using symbols. Fit to data using Eqs.~(\ref{R0}), (\ref{R90}) (c) and Eqs.~(\ref{T0}), (\ref{T90}) (d) are shown by the solid lines.
    Data shown were obtained with a device with $N = 1.46 \times 10^5$ and $U = 10.4$ mJ/cm$^2$.
    }
    \label{fig:angledep}
    \end{center}
\end{figure}

\subsection{Determination of the refractive index}

\subsubsection{Reflectance and transmittance data}

Exemplary data obtained using the reflection and transmission geometries are shown in Figs.~\ref{fig:angledep}(a) and  \ref{fig:angledep}(b), respectively, for the Te film, sapphire substrate and air with $\alpha = 0^\circ$ and $90^\circ$.
All data show a sinusoidal dependence on $\theta_\mathrm{WG}$.
To extract parameters from the data, we process the data as follows.
For reflection geometry, we subtract $V_{\mathrm{air},\alpha}^r$ from $V_{\mathrm{Te},\alpha}^r$ and normalize it with $(V_{\mathrm{s},h}^{r} - V_{\mathrm{air},h}^{r} + V_{\mathrm{s},v}^{r} - V_{\mathrm{air},v}^{r})$.
This procedure is used to suppress the background contribution arising from reflections on the lens surface, which is represented in $V_{\mathrm{air},\alpha}^r$.
The normalization also enables us to quantify the reflectance relative to the substrate independently of the incident laser power.
In the transmission geometry, $V_{\mathrm{Te},\alpha}^t$ is normalized by $(V_{\mathrm{air},h}^{t} + V_{\mathrm{air},v}^{t})$.
Here, the normalization enables us to quantify the transmittance relative to air, while removing the influence of the incident laser power.

Normalized voltages $V^t_{N,\alpha} = (V_{\mathrm{Te},\alpha}^r - V_{\mathrm{air},\alpha}^r)/(V_{\mathrm{s},h}^{r} - V_{\mathrm{air},h}^{r} + V_{\mathrm{s},v}^{r} - V_{\mathrm{air},v}^{r}), V^r_{N,\alpha} = V_{\mathrm{Te},\alpha}^t/(V_{\mathrm{air},h}^{t} + V_{\mathrm{air},v}^{t})$ are shown in Fig.~\ref{fig:angledep}(c) and \ref{fig:angledep}(d).
We fit these data to obtain the complex reflection and transmission coefficients of the sample.
$r_o$ and $r_e$ ($t_o$ and $t_e$) are defined as the reflection (transmission) coefficients when the probe laser polarization is parallel and orthogonal to pulse laser polarization, i.e. when the former is orthogonal and parallel to the expected $c$ axis of Te, respectively.
For the measurements of the sapphire substrate ($V_{\mathrm{s},\alpha}^{r/t}$), we introduce the complex reflection coefficients ($r_s$) and transmission coefficients ($t_s$) that represent the optical characteristics of the substrate.
Forms for the normalized voltages $V^t_{N,\alpha}$, $V^r_{N,\alpha}$ are obtained using the Jones formalism \cite{jones1941new} and presented in the Appendix, Sec.~\ref{sec:app:refl}.
See Eqs.~(\ref{R0}), (\ref{R90}) for the reflection geometry, Eqs.~(\ref{T0}), (\ref{T90}) for the transmission geometry and Eqs.~(\ref{s0}), (\ref{s90}) and (\ref{st}) for the sapphire substrate.
The transmission and reflection data are fitted together to extract the complex transmission and reflection coefficients.
With such simultaneous fitting, the number of fitting parameters can be reduced. 
Here we set $r_{o(e)} = |r_{o(e)}| \exp(i \phi_{o(e)}^r)$ and $t_{o(e)} = |t_{o(e)}| \exp(i \phi_{o(e)}^t)$, and define $\Delta \phi_t \equiv \mathrm{arg}(t_o/t_e) = \phi_{o}^t-\phi_{e}^t$ and $\Delta \phi_r \equiv \mathrm{arg} (r_o/r_e)=\phi_{o}^r-\phi_{e}^r$.
For the transmission (reflection) geometry, the fitting parameters are $|t_o|$, $|t_e|$ and $\Delta \phi_t$ ($|r_o|/|r_s|$, $|r_e|/|r_s|$ and $\Delta \phi_r$).
The fitting results are presented in Figs.~\ref{fig:angledep}(c) and  \ref{fig:angledep}(d) by the solid lines.
From the fitting shown in Figs.~\ref{fig:angledep}(c) and  \ref{fig:angledep}(d), we obtain $|t_o| = 0.502, |t_e| = 0.621,\Delta \phi_t = 13.7^\circ, |r_o|/|r_s| = 1.25, |r_e|/|r_s| = 1.13,\Delta \phi_r = -14.7^\circ$.
We performed a similar analysis on the substrate using the transmitted light and obtain $|t_s| = 0.873$, which also gives $|r_s| = \sqrt{1 - |t_s|^2} = 0.358$. These values are used to determine $|r_o|$ and $|r_e|$ for each sample.
The same analyses are performed for all samples. 

\begin{figure}[ht]
    \begin{center}
    \includegraphics[width=0.8\linewidth]{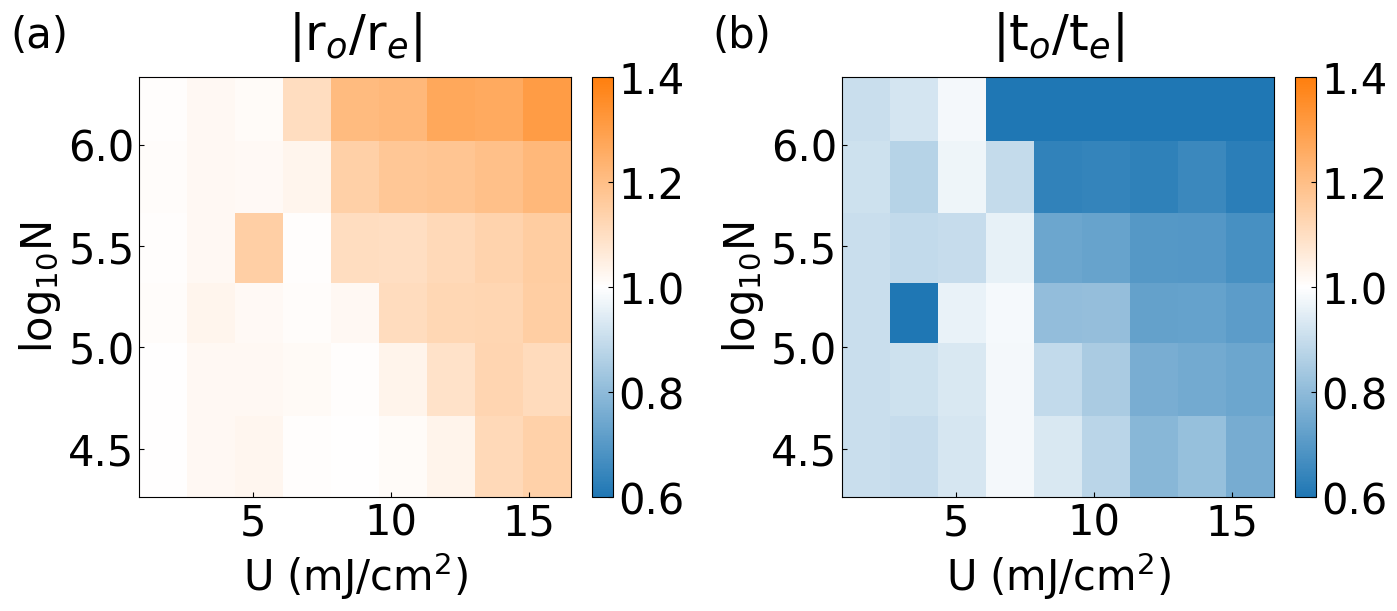}
    \caption{(a,b) Magnitude of the ratio of the reflection coefficient $\frac{|r_o|}{|r_e|}$ (a) and the transmission coefficient $\frac{|t_o|}{|t_e|}$ (b) plotted against laser fluence $U$ and pulse number $N$. The vertical axis is in log-scale.
    }
    \label{fig:transrefl}
    \end{center}
\end{figure}

In Fig.~\ref{fig:transrefl}, we show the dependence of magnitudes of $|r_o/r_e|$ and $|t_o/t_e|$ on the pulse fluence $U$ and pulse number $N$.
The color scale shows the ratios $\left| r_o/r_e \right|$ and $\left|t_o/t_e \right|$ and reflect the strength of the optical anisotropy. 
$\left|t_o/t_e \right|$ increases while $\left| r_o/r_e \right|$ decreases with increasing $U$ and $N$.
These trends indicate that higher-fluence irradiation with a larger number of pulses promotes Te crystal reorientation and enhances the optical anisotropy of the films.

\subsubsection{Refractive index of as deposited Te film}
From the complex transmission and reflection coefficients, we compute the refractive indices of the Te thin film irradiated with the pulse laser.
Let us assume that the index $i$ ($i=1,2,3$) indicates air, Te film, and substrate, respectively.
First, we calculate the refractive index of Te films that have not been irradiated with the pulse laser, i.e. the as deposited films.
Without laser irradiation, the Te $c$ axis is randomly oriented within the film plane\cite{kobayashi2026nanolett}.
We therefore assume $t_e = t_o \equiv t$ (thus $\Delta_t = 0$) and $r_e = r_o \equiv r$ (thus $\Delta_r = 0$). 
Taking into account the effect of light interference within the film \cite{azzam1978ellipsometry}, $t$ and $r$ read
\begin{gather}
\label{eq:rt}
r = \frac{r_{12} + r_{23}\exp(2i\Delta_2)}{1 + r_{12}r_{23}\exp(2i\Delta_2)},\ \ t = \frac{t_{12}t_{23}t_{31}\exp(i\Delta_2)}{1 + r_{12}r_{23}\exp(2i\Delta_2)},
\end{gather}
where $r_{ij}$ and $t_{ij}$ are the reflection and transmission amplitude at the media $i$ and $j$ interface:
\begin{gather}
\label{eq:rij:tij}
r_{ij}=\frac{\eta_i-\eta_j}{\eta_i+\eta_j},\ \ t_{ij} = \frac{2\eta_i}{\eta_i+\eta_j},
\end{gather}
and $\Delta_2$ is given by
\begin{gather}
\label{eq:Delta}
\Delta_2 = \frac{2\pi\eta_2 d}{\lambda}.
\end{gather}
$\eta_i$ is the refractive index for medium $i$ and we set $\eta_2 = n_2 + i k_2$ for Te and assume $\eta_1 = 1$ for air and $\eta_3 = 1.77$ for the sapphire (Al$_2$O$_3$) substrate.
The substrate was treated as an optically thick incoherent medium. 
Therefore, coherent Fabry--Perot interference associated with multiple internal reflections in the substrate was neglected \cite{katsidis2002general}.

For each coefficient ($r$, $t$), we set a cost function equal to the square of the difference between the calculated and experimentally obtained values divided by the experimental error.
We look for a set of variables ($n_2$, $k_2$, $d$) that minimize the sum of the cost functions.
From the process, we find $n_2 = 2.88$, $k_2 = 2.72$, and $d = 39~\mathrm{nm}$.
$n_2$ and $k_2$ are in good agreement with the literature\cite{guo2022nanoscale}: the average value of the ordinary ($n_o$, $k_o$) and extraordinary ($n_e$, $k_e$) components are $n_\mathrm{av} = \frac{1}{2} (n_o + n_e) = 2.80$, $k_\mathrm{av} = \frac{1}{2} (k_o + k_e) = 3.50$.
The thickness $d$ is consistent with the nominal value of $40~\mathrm{nm}$.
In the following, we consider $n_2$ and $k_2$ to be the average refractive index of Te and use these values in the following analyzes.

\subsubsection{\label{sec:refractiveindex:pulseTe} Refractive index of Te film irradiated with laser pulses}
Next, we calculate the refractive index of the Te films irradiated with pulse laser.
The refractive index of the Te film is now set to $\eta_{2,o} = n_o + i k_o$ ($\eta_{2,e} = n_e + i k_e$) when the incident light polarization is parallel (orthogonal) to the pulse laser polarization, i.e. orthogonal (parallel) to the expected Te crystal axis $c$ orientation. 
Again, we assume $\eta_1 = 1$ for air and $\eta_3 = 1.77$ for the sapphire substrate.
As previously reported \cite{kobayashi2026nanolett}, the surface roughness of the film increases with irradiation.
To account for the roughness, we model the roughness-induced attenuation of specular transmittance and reflectance via Debye-Waller-like exponential factors \cite{lin2011scalar}.
The thickness of the film is assumed to vary around a mean value with standard deviation $\sigma$ that follows a Gaussian distribution.
Taking into account the film surface roughness, the polarization dependent transmission and reflection coefficients of the system read: 
\begin{gather}
r_{o(e)} = \frac{\tilde{r}_{12,o(e)} + r_{23,o(e)}\exp(2i\Delta_{o(e)})}{1 + \tilde{r}_{12,o(e)}r_{23,o(e)}\exp(2i\Delta_{o(e)})},\ \ t_{o(e)}  = \frac{\tilde{t}_{12,o(e)}t_{23,o(e)}t_{31}\exp(i\Delta_{o(e)})}{1 + \tilde{r}_{12,o(e)}r_{23,o(e)}\exp(2i\Delta_{o(e)})},
\end{gather}
where the complex transmission and reflection amplitudes are defined as
\begin{gather}
r_{12,o(e)}=\frac{\eta_1-\eta_{2,o(e)}}{\eta_1+\eta_{2,o(e)}}, \ 
t_{12,o(e)} = \frac{2\eta_1}{\eta_1+\eta_{2,o(e)}},\ 
r_{23,o(e)}=\frac{\eta_{2,o(e)}-\eta_3}{\eta_{2,o(e)}+\eta_3},\ 
t_{23,o(e)} = \frac{2\eta_3}{\eta_{2,o(e)}+\eta_{3}},\\
\tilde{r}_{12,o(e)} = r_{12,o(e)}\exp\left(-\frac{8\pi^2\sigma^2}{\lambda^2}\right),\ 
\tilde{t}_{12,o(e)} = t_{12,o(e)}\exp\left(-\frac{2\pi^2(1-n_{2,o(e)})^2\sigma^2}{\lambda^2}\right),
\end{gather}
and $\Delta_{o(e)}$ is given by
\begin{gather}
\label{eq:Deltac}
\Delta_{o(e)} = \frac{2\pi\eta_{2,o(e)} d}{\lambda}.
\end{gather}
$o(e)$ represents the incident light polarization: it is parallel (orthogonal) to the pulse laser polarization. 
Again, we define a cost function for each parameter ($|r_e|$, $|t_e|$, $|r_o|$, $|t_o|$, $\Delta \phi_r$, $\Delta \phi_t$) and look for a set of variables ($\delta n_2 = n_e - n_o$, $\delta k_2 = k_e - k_o$, $d$, and $\sigma$) that minimize their sum.
As noted above, we fix $n_2 = \frac{1}{2}(n_e + n_o)$ and $k_2 = \frac{1}{2}(k_e + k_o)$ at their average values, and vary $\delta n_2$ and $\delta k_2$ when looking for the optimum set of variables.
The analysis is conducted for all devices.

$\delta n_2$ and $\delta k_2$ obtained from minimization are plotted in Fig.~\ref{fig:nk} against $U$ and $N$. 
See Fig.~\ref{fig:d:sigma} in the Appendix, Sec.~\ref{sec:app:par:reft}, for the film thickness $d$ and the interfacial roughness $\sigma$ estimated from the fitting.
We find that the signs of $\delta n_2$ and $\delta k_2$ are consistent with those reported in a previous study\cite{guo2022nanoscale}. 
The maximum value of $\delta k_2$ is close to 3, while $\delta n_2$ is in the range of $-1$ to 0. 
The latter is consistent with the literature\cite{guo2022nanoscale}, whereas the former is somewhat larger. 
We attribute the discrepancy of the obtained $\delta k$ with that from Ref.~\cite{guo2022nanoscale} to the treatment of roughness in the model. 
In the present analysis, roughness is incorporated through an exponential attenuation factor\cite{lin2011scalar}, which provides a reasonable first-order description when roughness primarily reduces the specular optical response through diffusive scattering. 
However, laser irradiation may generate laser-induced periodic surface structures (LIPSS) and other nanoscale corrugations on the surface \cite{sipe1983laser,bonse2020laser}, such that the photo-modified films may contain not only wavelength-scale roughness but also subwavelength interfacial features.
A more complete analysis would likely require, in addition to the present treatment, an effective-medium description of the near-surface region as an air-Te mixed layer \cite{fung2019application}.
However, incorporating both effects simultaneously is technically difficult and beyond the scope of the present work.
Because the model reproduces the signs of both $\delta n_2$ and $\delta k_2$ and yields a physically reasonable range of $\delta n_2$, we regard Fig.~\ref{fig:nk} as providing an adequate effective description of the optical response of the pulse-laser-irradiated Te films.

\begin{figure}[h]
    \begin{center}
    \includegraphics[width=0.8\linewidth]{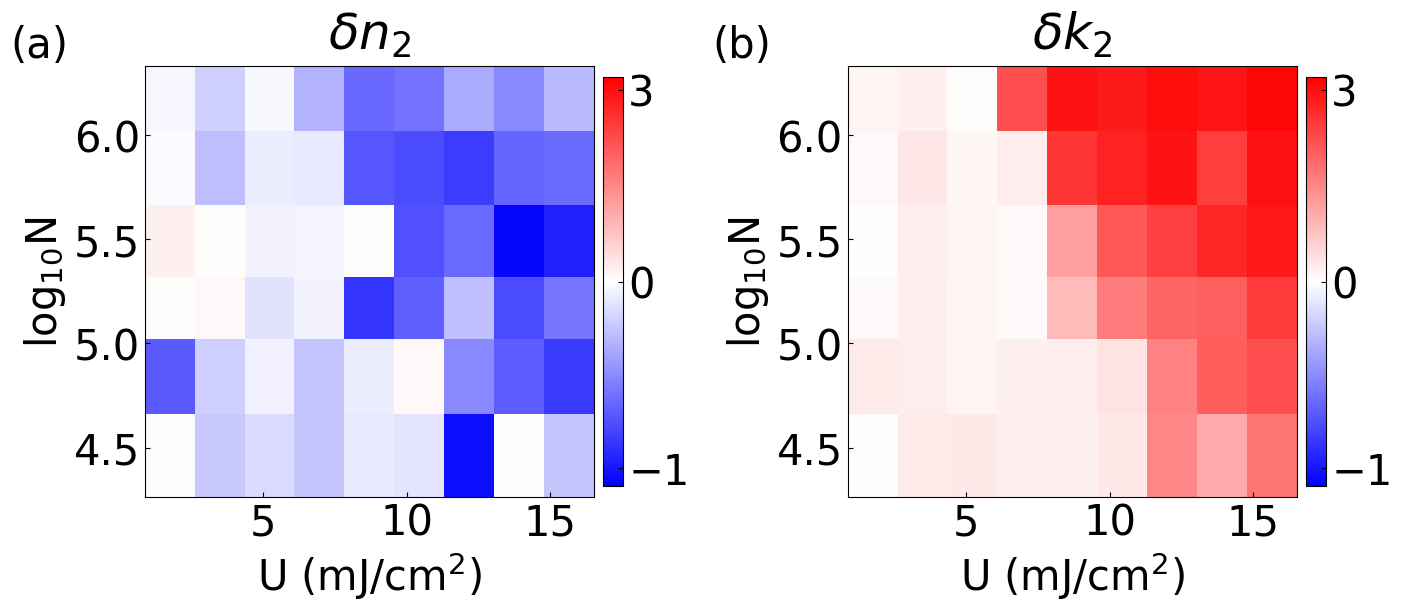}

    \caption{
    Laser fluence $U$ and pulse number $N$ dependence of the difference in the refractive indices $\delta n_2 = n_e - n_o$ (a) and extinction constants $\delta k_2 = k_e - k_o$ (b) when the Te $c$ axis points parallel ($n_o, k_o$) and perpendicular ($n_e, k_e$) to the incident laser polarization. The vertical axis is in log-scale.}
    \label{fig:nk}
    \end{center}
\end{figure}

\section{Thermally activated grain-reorientation model}

\subsection{Simple picture of laser-induced $c$ axis orientation}
Motivated by the description of all-optical magnetization switching (AOS), we construct a minimal kinetic model for laser irradiation-induced change in the $c$ axis reorientation implied by the results shown in Fig.~\ref{fig:nk}.
As is evident from Fig.~\ref{fig:nk}, $\delta n_2$ shows a weak dependence on $U$ and $N$, while $\delta k_2$ exhibits a sharp increase when $U$ or $N$ exceeds a certain threshold.
In all-optical switching (AOS)\cite{stanciu2007prl1,lambert2014science}, the switching probability has been reported to depend on both the laser fluence and the number of pulses \cite{khorsand2012prl,gorchon2016model,ellis2016all,hadri2016prb,yoshikawa2022cphys}. 
Such multishot AOS has been interpreted as a thermally activated process induced by magnetic circular dichroism.
By analogy, we consider that the laser-induced reorientation of Te is driven by anisotropic optical absorption arising from linear dichroism.

First, we describe the phenomenology of the minimal kinetic model.
When the Te film is irradiated with a linearly polarized pulse laser, grains whose $c$ axes are parallel to the light polarization absorb more optical energy than those oriented perpendicular to it because of the strong linear dichroism of Te.
The effective temperature of the former grains can reach a regime in which the orientation of the $c$ axis becomes reconfigurable.
After pulse excitation, the grains cool and the orientation of the $c$ axis is frozen to one of the metastable orientations.
In contrast, grains with lower optical absorption retain their original $c$ axis orientation as long as their effective temperature remains below the critical regime for reorientation.
After this process is repeated many times, an imbalance develops between the populations of grains with $c$ axes parallel and perpendicular to the pulse laser polarization.

\subsection{Model description}
For simplicity, let us assume that there are two metastable grain orientations: one with the $c$ axis parallel and the other perpendicular to the incident pulse laser polarization.
After irradiation of $n$ pulses, we define $P_{\parallel}(n)$ and $P_{\perp}(n)$ as the areal (or volumetric) fractions of grains with the $c$ axis parallel and perpendicular to the laser polarization, respectively, with $P_{\parallel}(n)+P_{\perp}(n)=1$.
We neglect thermal accumulation between successive laser pulses by assuming that each grain relaxes back to the initial temperature $T_0$ before the arrival of the next pulse.
The transient heating induced by each pulse is therefore represented by a step-like increase in the effective temperature, followed by a temperature drop over a characteristic relaxation time $\Delta t$.
Due to linear dichroism of Te, the effective temperature of the two metastable states differ.
We assume that the average temperature during the period of $\Delta t$ is $T_{\parallel}=T_0+\Delta T_{\parallel}$ and $T_{\perp}=T_0+\Delta T_{\perp}$ for the two states, which are defined as:
\begin{gather}
\Delta T_{\parallel} = \frac{2k_e}{\rho C d(k_e+ k_o)} A U ,\ \Delta T_{\perp} = \frac{2k_o}{\rho C d(k_e + k_o)} A U.
\label{eq:Ti}
\end{gather}
$k_i$ ($i=e,o$) is the extinction coefficient, $U$ is the incident laser fluence, $A$ is the absorbance of the film at the pulse-laser wavelength, $\rho$ is the density of Te, $C$ is the heat capacity, and $d$ is the film thickness.

During an elevated-temperature window of duration $\Delta t$ ($t=0$ corresponds to the time at which the laser pulse is applied), we assume the switching rate $\Gamma_{|| (\perp)}$ out of the initial state $\parallel$ ($\perp$) follows an Arrhenius law\cite{debye1929polar}:
\begin{equation}
\Gamma_{|| (\perp)}=
\frac{\nu_0}{2} \exp\!\left[-\dfrac{E V}{k_B T_{|| (\perp)}}\right],
\label{eq:rate}
\end{equation}
where $\parallel$ ($\perp$) represents the state with the $c$ axis parallel (perpendicular) to the incident pulse laser polarization.
$\nu_0$ is the attempt frequency, $E$ is the effective energy barrier density and $V$ is the activation volume.
We assume that $E$ is the same for both metastable states, regardless of the initial condition.
The factor of $1/2$ accounts for the fact that, after reaching the top of the energy barrier, the state is assumed to relax into one of the two metastable states, that is, the state can fall into the original state. 
This is in contrast to, e.g. magnetization switching, where the state switches to the other metastable state once the energy barrier is overcomed.
 
The population dynamics of the two metastable states during $\Delta t$ is governed by the master equation
\begin{gather}
    \frac{\dd}{\dd t}
\begin{pmatrix}
P_{\parallel}(n)\\
P_{\perp}(n)
\end{pmatrix}
=
\begin{pmatrix}
-\Gamma_\parallel&\Gamma_\perp\\
\Gamma_\parallel&-\Gamma_\perp
\end{pmatrix}
\begin{pmatrix}
P_{\parallel}(n)\\
P_{\perp}(n)
\end{pmatrix}.
\end{gather}
Using $P_{\parallel}(n)+P_{\perp}(n)=1$, the solution after the $n$th pulse is
\begin{equation}
\begin{aligned}
P_{\parallel}(n) = P_{\parallel}^{\mathrm{eq}} + \left[ P_{\parallel}(n-1) - P_{\parallel}^{\mathrm{eq}} \right] \exp \left[ -\left(\Gamma_\parallel + \Gamma_\perp \right) \Delta t \right],
\label{eq:Ppar}
\end{aligned}
\end{equation}
where $P_{\parallel}^{\mathrm{eq}}=\Gamma_\perp/(\Gamma_\parallel+\Gamma_\perp)$.
In deriving Eq.~(\ref{eq:Ppar}), we used the population after cooling from the $(n-1)$th pulse as the initial condition for the $n$th pulse:
$P_{\parallel}(n)|_{t=0}=P_{\parallel}(n-1)|_{t=\Delta t}$.
The areal fractions of the two grain orientations after irradiation by a total of $N$ pulses are obtained by iterating Eq.~(\ref{eq:Ppar}), producing
\begin{equation}
\begin{aligned}
P_{\parallel}(N) &= P_{\parallel}^{\mathrm{eq}} + \left[ P_{\parallel}(0) - P_{\parallel}^{\mathrm{eq}} \right] \exp \left[ -
\left(\Gamma_\parallel + \Gamma_\perp \right)N \Delta t \right],
\label{eq:Pperp:N}
\end{aligned}
\end{equation}
and
\begin{equation}
\begin{aligned}
P_{\perp}(N) &= 1 - P_{\parallel}(N).
\label{eq:Ppar:N}
\end{aligned}
\end{equation}

\subsection{Numerical estimation of the threshold energy}

In the following, we substitute the material and optical parameters in the above formulas to compute $P_{\perp}(N)$.
We assume that $E$ is the energy density required to heat Te from room temperature $T_0$ to the melting point $T_{\text{melt}}$, that is, 
\begin{equation}
\begin{aligned}
E = \rho C (T_{\text{melt}}-T_{0}) .
\label{eq:barrierheight}
\end{aligned}
\end{equation}
For Te, we use the following numbers: density $\rho=6240~\mathrm{kg\,m^{-3}}$\cite{thomson2001book}, specific heat capacity $C=0.202~\mathrm{J\,g^{-1}\,K^{-1}}$\cite{thomson2001book}
and melting point $T_{\mathrm{melt}}=722.66 \mathrm{K}$\cite{thomson2001book}.
The room temperature is set to $T_0=298.15\mathrm{K}$.
Substituting these parameters into Eq.~(\ref{eq:barrierheight}), we find $E = 5.35 \times 10^8~\mathrm{J\,m^{-3}}$.

The activation volume $V$ is assumed to be represented by an atomic-scale volume containing only a few atoms, which is based on the following three physical approximations. 
First, we consider the laser-induced reorientation to proceed through rearrangement of local chainlike structural units, rather than through rigid rotation of an entire crystalline grain.
This assumption is motivated by the chainlike crystal structure of Te, in which the relevant orientational degree of freedom corresponds to the direction of the local helical chain, namely the Te $c$ axis.
Accordingly, the activation volume $V$ should be regarded as a phenomenological volume associated with such a local rearrangement, not as the physical volume of a crystalline grain.
Second, we neglect the explicit dependence of $V$ on the lateral domain size and assume that the probability of local reorientation is governed primarily by the relative orientation between the light polarization and the Te $c$ axis through polarization-dependent absorption.
In an MBE-grown Te film, the crystallographic orientation may appear macroscopically random, although microscopic regions with partial orientational order of the helical chains may still exist.
The centers and edges of such domains may differ in thermal transport and optical absorption because grain boundaries can modify both heat flow and optical scattering or absorption\cite{smith2018grain,veprek1987effect}. 
In addition, finite domain size may lead to a reduction of the effective melting temperature through the domain size effect\cite{sun2007melting}.
A complete treatment of these effects would require a microscopic model of domain morphology, heat transport, and local structural rearrangement. 
Here, we instead adopt a coarse-grained description in which the influence of domain boundaries, neighboring crystallites, and finite-size effects is absorbed into the effective activation volume $V$.
Third, interchain correlations are not treated explicitly. 
Previous studies have shown that liquid Te near the melting point retains chainlike local motifs, while substantial interchain correlation is also present\cite{tsuzuki1995static}.
Therefore, even in the laser-heated or partially molten state, Te chains are not necessarily free to rotate independently.
In the present model, the effect of interchain correlation is incorporated phenomenologically into the activation volume $V$.

To estimate the order of magnitude of $V$, we use the average chain length reported for amorphous Te, 3.91 bonds excluding isolated atoms\cite{akola2012prb}, as a proxy for the size of the local structural unit involved in reorientation.
Combining this value with the atomic volume of trigonal Te yields $V \approx 1.67\times10^{-28}~\mathrm{m^3}$.
This value should therefore be regarded as an order-of-magnitude estimate of the effective reorienting volume, rather than as the exact geometrical volume of a molten grain.

The attempt frequency $\nu_0$ is set equal to $1/\Delta t$.
This is because $\Delta t$, the duration which the temperature is at or above $T_{\mathrm{melt}}$, is assumed to be comparable to the characteristic time window over which local structural rearrangement can occur in laser-heated Te.
See the discussion in Appendix, Sec.~\ref{sec:app:par:thermal}, for justification of such assumption.

The effective temperatures $T_{\parallel}$ and $T_{\perp}$ are obtained from Eq.~(\ref{eq:Ti}).
To evaluate Eq.~(\ref{eq:Ti}), the absorbance $A$ of the Te film at the wavelength of the pulse laser must be obtained.
$A$ is calculated from the reflectance $R$, transmittance $T$, and the relation $A = 1-R-T$.
The reflectance $R=|r|^2$ and transmittance $T=|t|^2$ are estimated using Eq.~(\ref{eq:rt}), the average complex refractive index $\eta_2$, and the film thickness $d$.
We use $\eta_2=\frac{1}{2}(\eta_{2,e}+\eta_{2,o})=4.62+1.20i$ at $\lambda=1030$ nm from Ref.~\cite{ciesielski2018matscisemiproc}.
Although the polarization-dependent refractive index in the visible range was reported in \cite{guo2022nanoscale}, we are not aware of the value at $\lambda=1030$ nm.
Here, we take the value from Ref.~\cite{guo2022nanoscale} at the longest wavelength (900 nm): $\eta_{2,e}-\eta_{2,o}=0.1369+1.2161i$.
From $\eta_2$ and $\eta_{2,e}-\eta_{2,o}$, we extract the extinction coefficients $k_i$ ($i=e,o$), which are substituted in Eq.~(\ref{eq:Ti}).
The film thickness is set to $d=40$ nm.

\begin{figure}[ht]
    \begin{center}
    \includegraphics[width=0.8\linewidth]{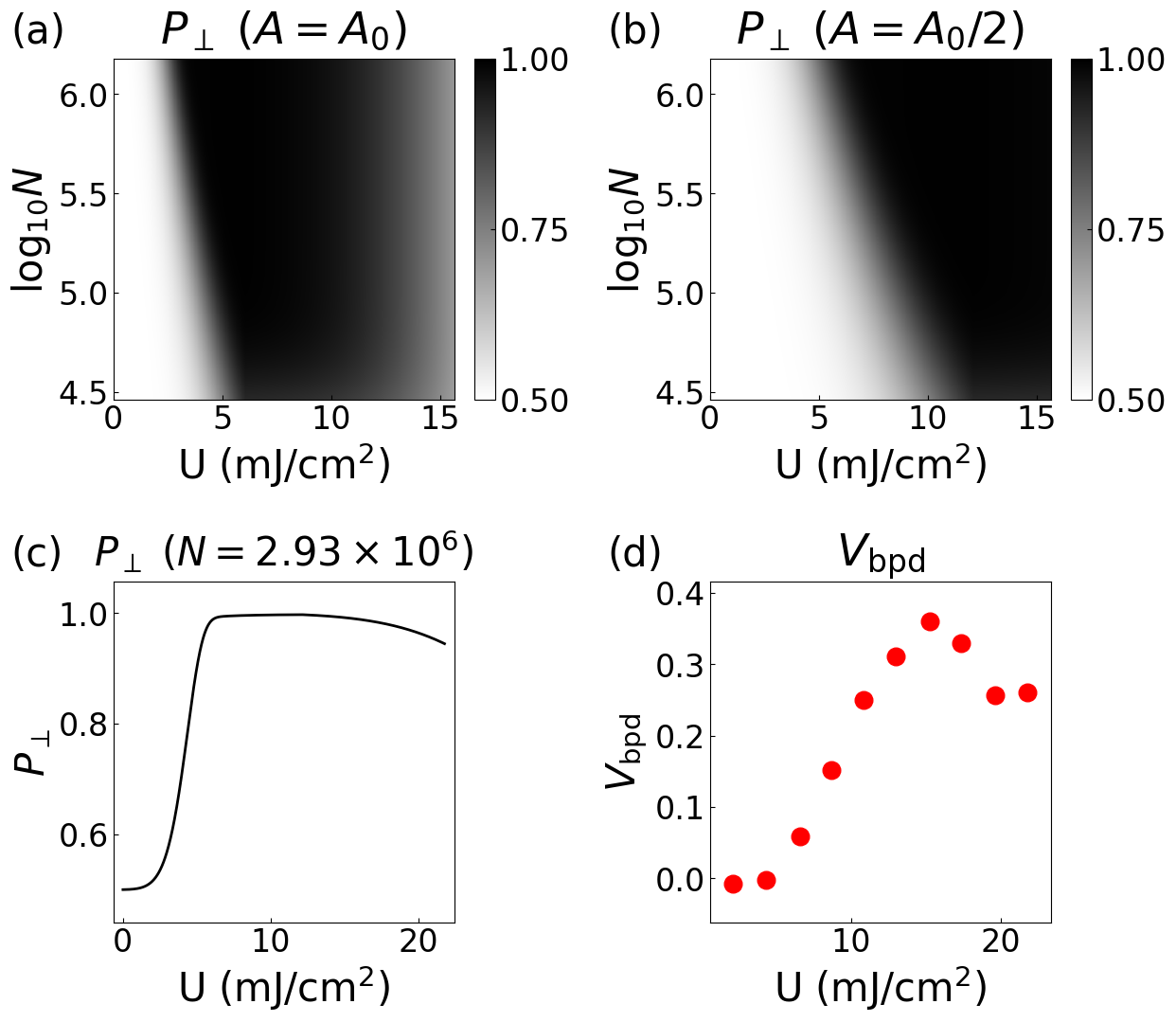}
\caption{\label{fig:pperp_modelcalc} (a) Calculated areal fraction $P_\perp$ of grains with the $c$ axis perpendicular to the pulse-laser polarization plotted as a function of laser fluence $U$ and the number of pulses $N$. Equation~(\ref{eq:Ppar:N}) is used to calculate $P_\perp$.
(b) Calculated areal fraction $P_\perp$ with the absorbance $A$ set equal to half of that used in (a).
(c) Line profile of $P_\perp$ from 
(b)
with $N=2.93\times10^6$.
(d) $V_\mathrm{bpd}$, proportional to $P_\perp$, plotted against $U$, reproduced from Ref.~\cite{kobayashi2026nanolett}. The data is obtained with $N=2.93\times10^6$.
}
    \end{center}
\end{figure}

Before irradiation, the grains in the film are assumed to be randomly oriented, resulting in an initial population $P_{\parallel}(0)=1/2$.
These values, along with fluence $U$, are substituted in Eqs.~(\ref{eq:Ti}) - (\ref{eq:rate}) and Eqs.~(\ref{eq:Pperp:N}) -- (\ref{eq:Ppar:N}).
The calculated areal fraction of grains with the $c$ axis perpendicular to the pulse-laser polarization is shown in Fig.~\ref{fig:pperp_modelcalc}.
The $U$ and $N$ dependence of $P_\perp$ shows features similar to those of $\delta k$ presented in Fig.~\ref{fig:nk}(b) in the low-fluence region ($U<10~\mathrm{mJ\,cm^{-2}}$).
In particular, $P_\perp$ exhibits an $N$-dependent threshold as a function of $U$.
These results support the interpretation that anisotropic laser heating biases thermally activated local reorientation and thereby aligns the crystallographic orientation relative to the pulse-laser polarization.

\section{Discussion}
\subsection{Comparison of the model calculation results with experiments}
The calculated threshold $U$ is lower than that obtained from the experiments, which we attribute primarily to an overestimation of the temperature rise.
In the present model, we employ a crude approximation in which thermal diffusion is neglected during the time interval $\Delta t$.
In reality, heat diffusion occurs concurrently with laser-induced heating; therefore, the peak temperature can be overestimated.
Such an overestimation of the heating efficiency can be corrected phenomenologically by reducing the optical absorbance $A$ of the thin film.
Figure~\ref{fig:pperp_modelcalc}(a) shows the result calculated using the absorbance $A_0$ which was estimated from the film thickness $d =40$ nm and the average complex refractive index reported in previous studies \cite{ciesielski2018matscisemiproc}, whereas Figure~\ref{fig:pperp_modelcalc}(b) shows the result calculated with $A=A_0/2$.
The trend obtained for $A=A_0/2$ reproduces the experimental trend more closely.
The calculations also suggest a high-$U$ ($U > 10$ mJ cm$^{-2}$) regime in which $P_{\perp}$ decreases.
Figure~\ref{fig:pperp_modelcalc}(c) shows the calculated $P_{\perp}$ obtained under the $N$ used in the previous study.
As a reference, the fluence dependence of $V_\mathrm{bpd}$, which is proportional to $P_{\perp}$, is shown in Figure~\ref{fig:pperp_modelcalc}(d).
In both cases, \(P_{\perp}\) initially increases with \(U\) and then decreases in the high-\(U\) regime, consistent with Ref.~\cite{kobayashi2026nanolett}.
We attribute this behavior to the onset of orientation-independent melting of Te grains at sufficiently high laser fluence.

\subsection{Assessment of other anisotropic materials}

Our model is governed primarily by anisotropic optical absorption in thin films and can therefore be extended to other anisotropic materials.
For this purpose, we rewrite the parameters entering the transition rates in terms of three material-dependent quantities, $x$, $y$, and $z$, and apply the formulation to optically anisotropic van der Waals materials BP, WTe$_2$, and SnSe.
These parameters are defined as follows:
\begin{equation}
\begin{aligned}
x = \frac{EV}{k_B T_0}, y = \frac{k_e - k_o}{k_e + k_o} , z = \frac{A}{\rho CdT_0},
\label{eq:xyz}
\end{aligned}
\end{equation}
where $x$ represents the reaction-barrier height, $y$ characterizes the anisotropy of the extinction coefficients, and $z$ quantifies the efficiency of the temperature rise.
With Eqs.~(\ref{eq:Ti}) and (\ref{eq:rate}), the transition rates can be rewritten as
\begin{equation}
\begin{aligned}
\Gamma_{\perp} &= f(U,x,1-y,z),\\
\Gamma_{\parallel} &= f(U,x,1+y,z),\\
&f(U,X,Y,Z) = \frac{\nu_0}{2} \exp\left[-\frac{X}{1 + Y Z U}\right],\\
\end{aligned}
\label{eq:Gamma:f}
\end{equation}

In these layered materials, we assume that the relevant activation volume is set by the smallest structurally meaningful unit that can be locally reconfigured during laser-induced heating, namely a unit-cell-sized portion of a single structural layer or sheet, rather than by a three-dimensional bulk-like grain.
For SnSe and WTe$_2$, previous studies have discussed thermally activated vacancy formation, vacancy clustering, and chalcogen-related point defects in layered chalcogenides.\cite{sraitrova2019vacancies,song2018single}
These results indicate that, upon heating, local atomic configurations can be modified through the formation, migration, or aggregation of vacancies, particularly at chalcogen sites.
Although atomic motion under pulsed-laser irradiation depends on the material and irradiation conditions, transient heating can promote local bond rearrangement and defect-assisted atomic motion before heat is dissipated into the substrate.\cite{kang2021phase}
Thus, in applying the present kinetic model to these materials, the activation process should be interpreted as a local atomic-scale rearrangement rather than as the collective reorientation of an entire grain.
Similarly, black phosphorus consists of puckered covalent phosphorus layers stacked through van der Waals interactions.\cite{appalakondaiah2012effect}
If a comparable laser-induced local rearrangement occurs in BP, $V$ should also be determined by the local structural volume similar to that assumed for Te.
\begin{figure}[tb]
    \begin{center}
    \includegraphics[width=0.6\linewidth]{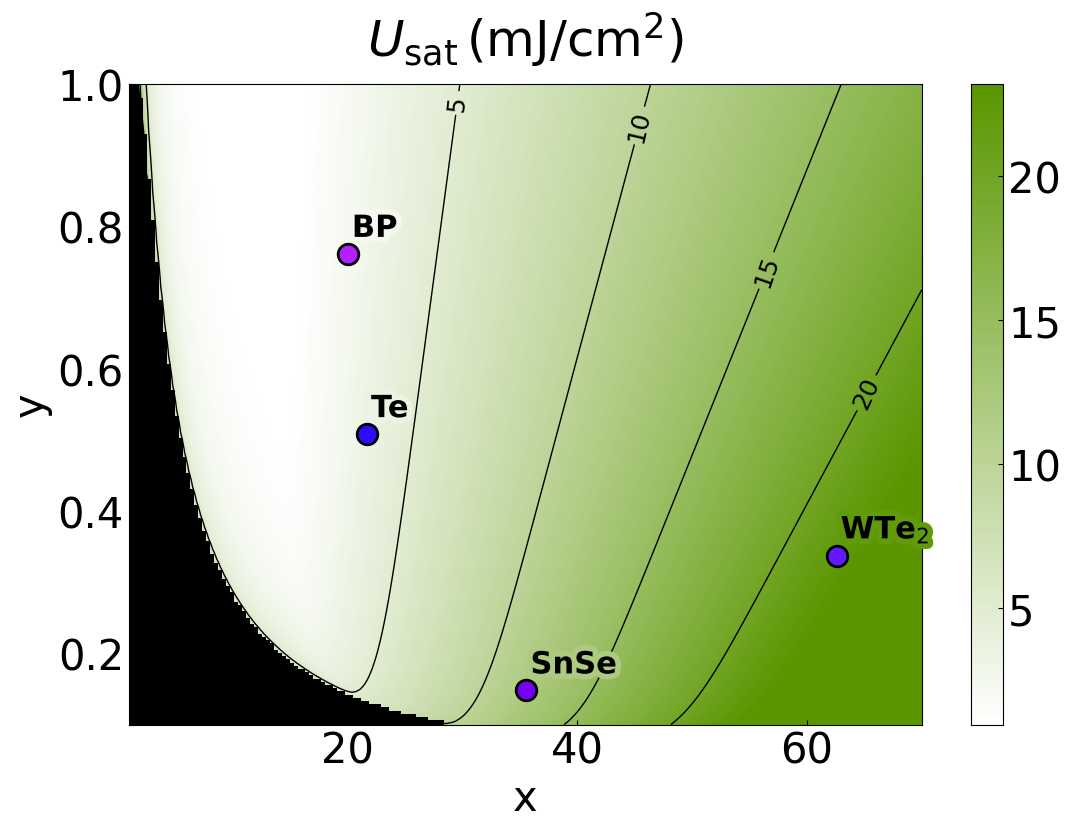}
    \caption{Calculated $U_{\mathrm{sat}}$ plotted as a function of scaled material parameters $x$ and $y$. The corresponding positions of Te, BP, WTe$_2$, and SnSe in the $(x,y)$ parameter space are indicated by circles. The black region indicates laser induced $c$ axis orientation with 90\% grain alignment is not possible.
    }
    \label{fig:Usat}
    \end{center}
\end{figure}

The effective layer thickness of these layered materials (SnSe,WTe$_2$,BP) was approximated as one half of the bulk lattice constant along the layer-normal direction, corresponding to the case where the bulk unit cell contains two equivalent layers or sheets.
Accordingly, the activation volume $V$ was assumed, as a first approximation, to be equal to one half of the unit-cell volume.
This treatment neglects possible cooperative rearrangements extending over multiple layers and should therefore be regarded as an order-of-magnitude estimate.

We calculate the critical laser fluence required to control crystal structure of the anisotropic materials.
Among Te, BP, WTe$_2$ and SnSe, the parameter $z$ has similar values, $z \approx 0.1~\mathrm{cm^2\,mJ^{-1}}$.
We therefore fix $z$ to this value and set $N = 10^6$.
Using this parameterization, we calculate the fluence $U_{\mathrm{sat}}$, defined as the value of $U$ at which $P_{\perp} = 90$\% is achieved.
To estimate $x$ and $y$, we used the reported values of the heat capacity \cite{Stephenson1969BPHeatCapacity,Callanan1992WTe2HeatCapacity,Wiedemeier1981SnSeHeatCapacity}, optical constants \cite{Lee2022BPPhaseRetarder,Munkhbat2022TMDOpticalConstants,Guo2023SnSePolarizer}, lattice constants \cite{Brown1965BPStructure,Brown1966WTe2Structure,Chattopadhyay1986SnSeStructure}, and melting points \cite{muhammad2024melting,Hansen2020WTe2PeritecticMelting,feutelais1996phase}. 
The parameters used are summarized in the Appendix, Sec.~\ref{sec:app:materials}.

The dependence $(x,y)$ of $U_{\mathrm{sat}}$ are plotted in Fig.~\ref{fig:Usat}.
The results shown in Fig.~\ref{fig:Usat} indicate that BP, WTe$_2$, and SnSe could also be structurally manipulated by linearly polarized pulsed-laser irradiation at fluences of the same order as those used in Te.

\section{Conclusion}
In summary, we have studied pulse laser-induced reorientation of the crystallographic $c$ axis in tellurium thin films.
The $c$ axis is aligned perpendicular to the polarization of the pulse laser.
The laser fluence $U$ and the number of pulses $N$ are varied to study the mechanism by which the $c$ axis alignment occurs.
We find that probability of the alignment increases systematically with peak power and pulse number. 
To account for these results, we develop a model that considers linear dichorism of Te and anisotropic heating and relaxation as the cause of laser-induced $c$ axis orientation. 
The model reproduces the observed trends, supporting a photothermal origin of the alignment.
We also estimate the laser fluence required to set the optic axis orientation of other anisotropic materials, e.g. WTe$_2$, SnSe and BP.
Substituting relevant parameters, we find the required fluence to align the optic axis is similar to that of Te. 
These results therefore demonstrate a route to anisotropic photoprocessing of low-dimensional materials, with implications for reconfigurable optoelectronics (including active metasurfaces) and anisotropic spintronic devices.

\section*{Acknowledgments}
The authors thank Kuniaki Konishi and Haruyuki Sakurai for fruitful discussions on the optical measurements and analyses. This work was partly supported by JSPS KAKENHI (Grant Numbers 23H00176,25K22216), JST CREST (JPMJCR19T3), MEXT Initiative to Establish Next-generation Novel Integrated Circuits Centers (X-NICS) and Cooperative Research Project Program of RIEC, Tohoku University. A.M. is supported by Materials Education program for the future leaders in Research, Industry, and Technology (MERIT), The University of Tokyo. T.H is supported by the Forefront Physics and Mathematics Program to Drive Transformation (FoPM), The University of Tokyo.

\clearpage

\section{Appendix}
\subsection{\label{sec:app:refl} Derivation of the forms used in the reflectance, transmittance measurements}
Here we derive the forms of $V_{\mathrm{Te},\alpha}^{r/t}$, $V_{\mathrm{s},\alpha}^{r/t}$, and $V_{\mathrm{air},\alpha}^{r/t}$, which are used to fit the experimental data presented (e.g., Fig.~\ref{fig:angledep}).
We use the Jones formalism to derive the following relations.
The final results are shown in Eqs.~(\ref{T0})-(\ref{R90}).

The polarization state of the light right before the sample $E$ can be described in the Jones formalism as
\begin{align}
	\label{eq:jones:E}
	E = M_\mathrm{GTP(T/R)} M_\mathrm{Te}^\mathrm{t/r} M_\mathrm{WG} M_\mathrm{QWP}(45^\circ) M_\mathrm{GTP} E_\mathrm{laser},
\end{align}
where $M_\mathrm{GTP(T/R)} = M_\mathrm{P}(\alpha)$, $M_\mathrm{WG}= M_\mathrm{P}(\theta_\mathrm{WG})$ and $M_\mathrm{GTP} = M_\mathrm{P}(0)$ are the Jones matrices representing the GTP(T / R), WG, and GTP, respectively.
$M_\mathrm{P}(\varphi)$ is the Jones matrix for a polarizer with its fast axis rotated $\varphi$ from the vertical axis:
\begin{align}
	\label{eq:jones:pol}
	M_\mathrm{P}(\varphi) = 
	\left[
        \begin{matrix} 
		\cos^2 \varphi & \cos \varphi \sin \varphi\\
		\cos \varphi \sin \varphi & \sin^2 \varphi
        \end{matrix}
        \right], 
\end{align}
Similarly, $M_\mathrm{QWP}(\varphi)$ is the Jones matrix for a quarter wave plate, which reads
\begin{align}
	\label{eq:jones:qwp}
	M_\mathrm{QWP}(\varphi) = \frac{1}{2}
	\left[
        \begin{matrix} 
		 (1+ i) + (1 - i) \cos 2 \varphi & (1 - i) \sin 2 \varphi\\
		(1 - i) \sin 2 \varphi & (1+ i) - (1 - i) \cos 2 \varphi
        \end{matrix}
        \right], 
\end{align}
$E_\mathrm{laser}$ is the state of light after the laser. In the experiments, the polarization is set along the $y$-axis.
\begin{align}
	\label{eq:jones:laser}
	E_\mathrm{laser} = E_0
    	\left[
        \begin{matrix} 
		1\\
		0
        \end{matrix}
        \right], 
\end{align}
where $E_0$ is the amplitude of the laser electric field.
$M_\mathrm{Te}^\mathrm{t(r)}$ is the Jones matrix of Te, defined as
\begin{gather}
	\label{eq:MTe:T}
	M_\mathrm{Te}^t = 
    	\left[
        \begin{matrix} 
        		\cos\delta & -\sin\delta\\
		\sin\delta & \cos\delta
        \end{matrix}
        \right]
        \left[
        \begin{matrix} 
        		t_e & 0\\
		0 & t_o
        \end{matrix}
        \right]
        \left[
        \begin{matrix} 
        		\cos\delta & \sin\delta\\
		-\sin\delta & \cos\delta
        \end{matrix}
        \right],
\end{gather}
for the transmission geometry and 
\begin{gather}
	\label{eq:MTe:R}
	M_\mathrm{Te}^r = 
    	\left[
        \begin{matrix} 
        		\cos\delta & -\sin\delta\\
		\sin\delta & \cos\delta
        \end{matrix}
        \right]
        \left[
        \begin{matrix} 
        		r_e & 0\\
		0 & r_o
        \end{matrix}
        \right]
        \left[
        \begin{matrix} 
        		\cos\delta & \sin\delta\\
		-\sin\delta & \cos\delta
        \end{matrix}
        \right],
\end{gather}
for the reflection geometry.

Let us start from the transmission geometry. 
Setting $\alpha = 0^\circ$ for GTP(T), the light intensity detected by the PD is given by $E^\dagger \cdot E$, which reads
\begin{gather}
V_{\mathrm{Te}, v}^t 
=\frac{I}{2}[(|t_e|^2\cos^2\delta + |t_o|^2\sin^2\delta) + (|t_e|^2\cos^2\delta - |t_o|^2\sin^2\delta)\cos(2(\theta-\delta)) \notag\\-|t_e t_o|\cos\Delta \phi_t \sin2\delta \sin(2(\theta-\delta))]\label{t0},
\end{gather}
where $I = E_0^2$ .
Similarly, configuration with $\alpha = 90^\circ$ returns
\begin{gather}
V_{\mathrm{Te}, h}^t 
= \frac{I}{2}[(|t_e|^2\sin^2\delta + |t_o|^2\cos^2\delta) + (|t_e|^2\sin^2\delta - |t_o|^2\cos^2\delta)\cos(2(\theta-\delta)) \notag\\+|t_e t_o|\cos\Delta \phi_t \sin2\delta \sin(2(\theta-\delta))]\label{t90},
\end{gather}
In the absence of the sample, the corresponding signals are obtained by setting $t_e,t_o\rightarrow 1$, resulting in the following.
\begin{gather}
V_{\mathrm{air}, v}^t
=\frac{I}{2}[1+ \cos2\theta], \label{air0}\\
V_{\mathrm{air}, h}^t
=\frac{I}{2}[1- \cos2\theta], \label{air90}
\end{gather}
We therefore obtain the normalized voltage,$V^t_{N,v},V^t_{N,h}$.
\begin{gather}
    V^t_{N,v} \equiv
\frac{V_{\mathrm{Te}, v}^t}{V_{\mathrm{air}, v}^t + V_{\mathrm{air}, h}^t} \notag\\
= \frac{1}{2}[(|t_e|^2\cos^2\delta + |t_o|^2\sin^2\delta) + (|t_e|^2\cos^2\delta - |t_o|^2\sin^2\delta)\cos(2(\theta-\delta)) \notag\\
-|t_e t_o|\cos\Delta \phi_t \sin2\delta \sin(2(\theta-\delta))]\label{T0},\\
    V^t_{N,h} \equiv \frac{V_{\mathrm{Te}, h}^t}{V_{\mathrm{air}, v}^t + V_{\mathrm{air}, h}^t} \notag\\
= \frac{1}{2}[(|t_e|^2\sin^2\delta + |t_o|^2\cos^2\delta) + (|t_e|^2\sin^2\delta - |t_o|^2\cos^2\delta)\cos(2(\theta-\delta)) \notag\\
+|t_e t_o|\cos\Delta \phi_t \sin2\delta \sin(2(\theta-\delta))]\label{T90}.
\end{gather}

Next, we look for the forms under the reflection geometry.
When the sample is removed, the measured signal mainly originates from reflections on the surface of the objective lens.
The reflection signals can be written as
\begin{gather}
V_{\mathrm{Te}, v}^r- V_{\mathrm{air}, v}^r 
=\frac{I}{2}[(|r_e|^2\cos^2\delta + |r_o|^2\sin^2\delta) + (|r_e|^2\cos^2\delta - |r_o|^2\sin^2\delta)\cos(2(\theta-\delta)) \notag\\-|r_e r_o|\cos\Delta \phi_r \sin2\delta \sin(2(\theta-\delta))]\label{r0},\\
V_{\mathrm{Te}, h}^r- V_{\mathrm{air}, h}^r 
= \frac{I}{2}[(|r_e|^2\sin^2\delta + |r_o|^2\cos^2\delta) + (|r_e|^2\sin^2\delta - |r_o|^2\cos^2\delta)\cos(2(\theta-\delta)) \notag\\+|r_e r_o|\cos\Delta \phi_r \sin2\delta \sin(2(\theta-\delta))]\label{r90},
\end{gather}

As noted in the main text, the optical anisotropy of the sapphire substrate is negligible when the incident light is from the substrate surface normal.
We therefore assume that the Jones matrix of the substrate takes the following form:
\begin{gather}
	\label{eq:Ms:T}
	M_\mathrm{s}^t = 
    	\left[
        \begin{matrix} 
        		t_s & 0\\
		0 & t_s
        \end{matrix}
        \right],
\end{gather}
for the transmission geometry and 
\begin{gather}
	\label{eq:Ms:R}
	M_\mathrm{s}^r = 
    	\left[
        \begin{matrix} 
        		r_s & 0\\
		0 & r_s
        \end{matrix}
        \right],
\end{gather}
for the reflection geometry.
Changing $M_\mathrm{Te}^\mathrm{t(r)}$ with $M_\mathrm{s}^\mathrm{t(r)}$, we find the following.
\begin{gather}
V_{\mathrm{s}, v}^r- V_{\mathrm{air}, v}^r
=\frac{I|r_\mathrm{s}|^2}{2}[1+ \cos2\theta]\label{s0},\\
V_{\mathrm{s}, h}^r- V_{\mathrm{air}, h}^r
=\frac{I|r_\mathrm{s}|^2}{2}[1- \cos2\theta]\label{s90},
\end{gather}
From Eqs.~(\ref{r0}), (\ref{r90}), (\ref{s0}) and (\ref{s90}), we obtain the normalized voltage, $V^r_{N,v}, V^r_{N,h}$

\begin{gather}
    V^r_{N,v} \equiv \frac{V_{\mathrm{Te}, v}^r-V_{\mathrm{air}, v}^r}{V_{\mathrm{s}, v}^r-V_{\mathrm{air}, v}^r + V_{\mathrm{s}, h}^r-V_{\mathrm{air}, h}^r}\notag\\
= \frac{1}{2|r_\mathrm{s}|^2}[(|r_e|^2\cos^2\delta + |r_o|^2\sin^2\delta) + (|r_e|^2\cos^2\delta - |r_o|^2\sin^2\delta)\cos(2(\theta-\delta)) \notag\\
-|r_e r_o|\cos\Delta \phi_r \sin2\delta \sin(2(\theta-\delta))]\label{R0},\\
    V^r_{N,h} \equiv\frac{V_{\mathrm{Te}, h}^r-V_{\mathrm{air}, h}^r}{V_{\mathrm{s}, v}^r-V_{\mathrm{air}, v}^r + V_{\mathrm{s}, h}^r-V_{\mathrm{air}, h}^r}\notag\\
= \frac{1}{2|r_\mathrm{s}|^2}[(|r_e|^2\sin^2\delta + |r_o|^2\cos^2\delta) + (|r_e|^2\sin^2\delta - |r_o|^2\cos^2\delta)\cos(2(\theta-\delta)) \notag\\
+|r_e r_o|\cos\Delta \phi_r \sin2\delta \sin(2(\theta-\delta))]\label{R90}.
\end{gather}
for the reflection geometry with the Te film.
Under the transmission and reflection geometries when a sapphire substrate is measured, we find the following.
\begin{gather}
\frac{V_{\mathrm{s}, v}^t+V_{\mathrm{s}, h}^t}{V_{\mathrm{air}, v}^t + V_{\mathrm{air}, h}^t} = |t_\mathrm{s}|^2\label{st},\\
|r_\mathrm{s}|^2 = 1-|t_\mathrm{s}|^2.\label{sr}
\end{gather}

\subsection{\label{sec:app:par:reft} Fitting parameters from the reflectance, transmittance measurements}
\begin{figure}[ht]
  \centering
  \noindent
    \includegraphics[width=0.4\linewidth]{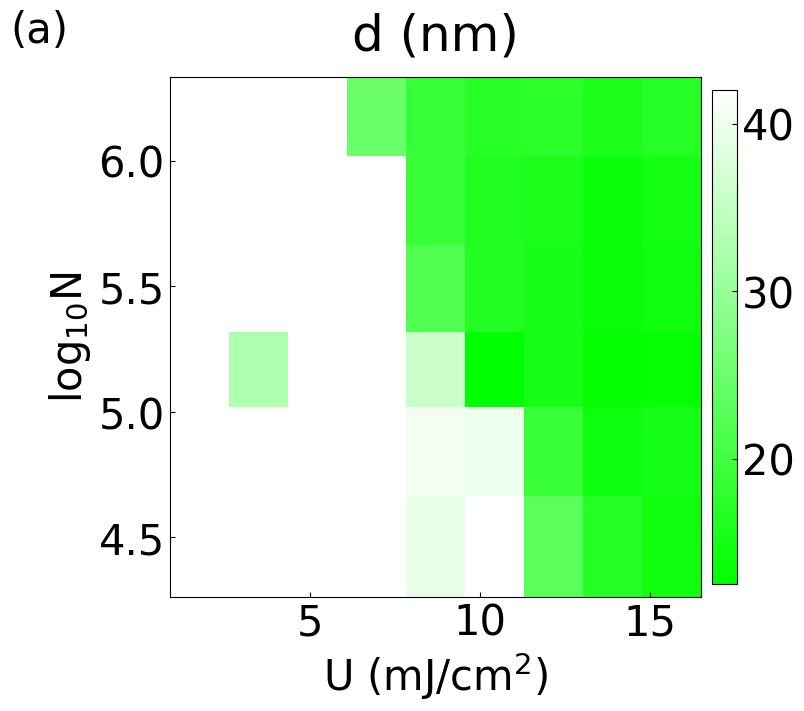}
    \includegraphics[width=0.4\linewidth]{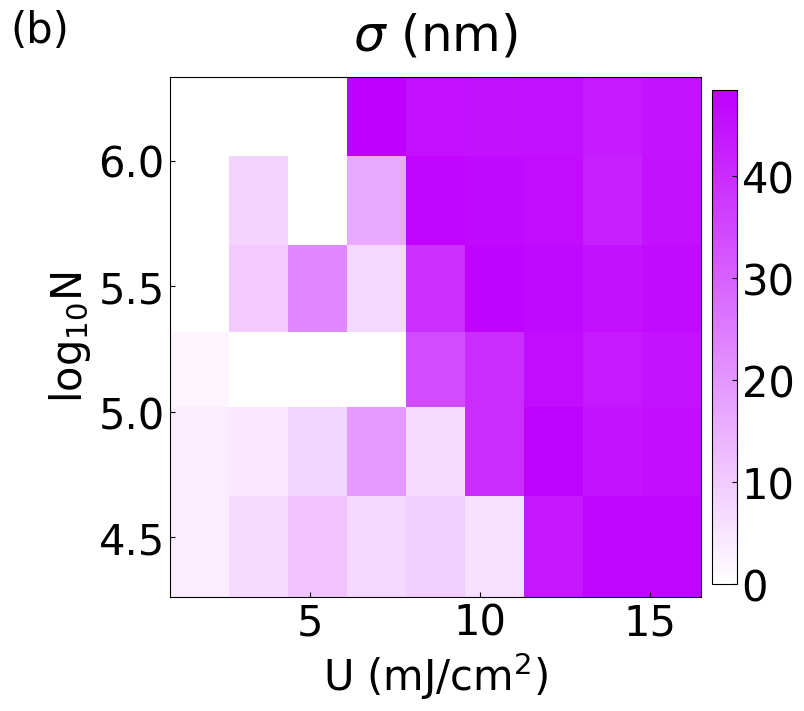}
\caption{Results Laser fluence $U$ and the number of pulses $N$ dependence of the estimated film thickness $d$ (a) and the corresponding film roughness $\sigma$ (b) when the Te $c$ axis points parallel ($n_o, k_o$) and perpendicular ($n_e, k_e$) to the incident laser polarization. The vertical axis is in log-scale. 
}
\label{fig:d:sigma}
\end{figure}

In this section, we show the parameters obtained by fitting the experimental data with the polarization dependent transmission and reflection model that includes surface roughness, discussed in Sect.~\ref{sec:refractiveindex:pulseTe}.
In Fig.~\ref{fig:nk}, we presented $\delta n_2$ and $\delta k_2$ as a function of $U$ and $N$.
Here, the other two fitting parameters, the film thickness $d$ and the interfacial roughness $\sigma$ are plotted against $U$ and $N$ in Fig.~\ref{fig:d:sigma}.
Comparing the results with those shown in Fig.~\ref{fig:nk}, we find that the region where the anisotropic response emerges is characterized by a reduced thickness and increased roughness.
This trend is consistent with partial laser-induced evaporation of Te accompanied by pronounced surface corrugation, suggesting that the observed reorientation is thermally driven.

\subsection{\label{sec:app:par:thermal} Time scale of laser induced heating}
We estimate the characteristic attempt frequency $\nu_0$ associated with thermally activated reorientation of Te grains by combining a one-dimensional heat-transport simulation with an order-of-magnitude estimate of the characteristic time scale over which rotational motion becomes randomized.

First, we calculate the heat transport in a Te thin film.
We consider the heat flow along the through-thickness (depth) direction, which reduces the problem to an 1 dimensional heat-transport problem. 
Assuming Beer-Lambert absorption 
\cite{chartier2005book}
with absorption length $l$, the temperature $T(x,t)$ of the system satisfies
\begin{gather}
\rho C \,\frac{\partial T}{\partial t}
= \kappa \,\frac{\partial^2 T}{\partial x^2} + S(x,t),\\
S(x,t)=\frac{I(t)}{l} e^{-x/l},\qquad
I(t)=\frac{U}{\sqrt{2\pi}\sigma}\exp\!\left[-\frac{(t-t_0)^2}{2\sigma^2}\right].
\end{gather}
We impose an adiabatic boundary condition at the free surface ($x=0$),
\begin{gather}
\left.\frac{\partial T}{\partial x}\right|_{x=0}=0,
\end{gather}
and a Robin boundary condition at the Te / Al$_{2}$O$_{3}$ interface ($x=d$),
\begin{gather}
-\kappa \left.\frac{\partial T}{\partial x}\right|_{x=d}
= G\,[T(d,t)-T_0],
\end{gather}
where $T_0=273~\mathrm{K}$ is the bath temperature.
Since the thermal boundary conductance of the Te / Al$_{2}$O$_{3}$ interface is not available in the
literature to the best of our knowledge, we use as an approximation the reported value for the
Bi / Al$_{2}$O$_{3}$ interface, $G=1950~\mathrm{W\,cm^{-2}\,K^{-1}}$ \cite{sheu2011kapitza}, noting that $G$ should be considered as an estimate of order-of-magnitude.

\begin{figure}[ht]
  \centering
  \includegraphics[width=0.6\linewidth]{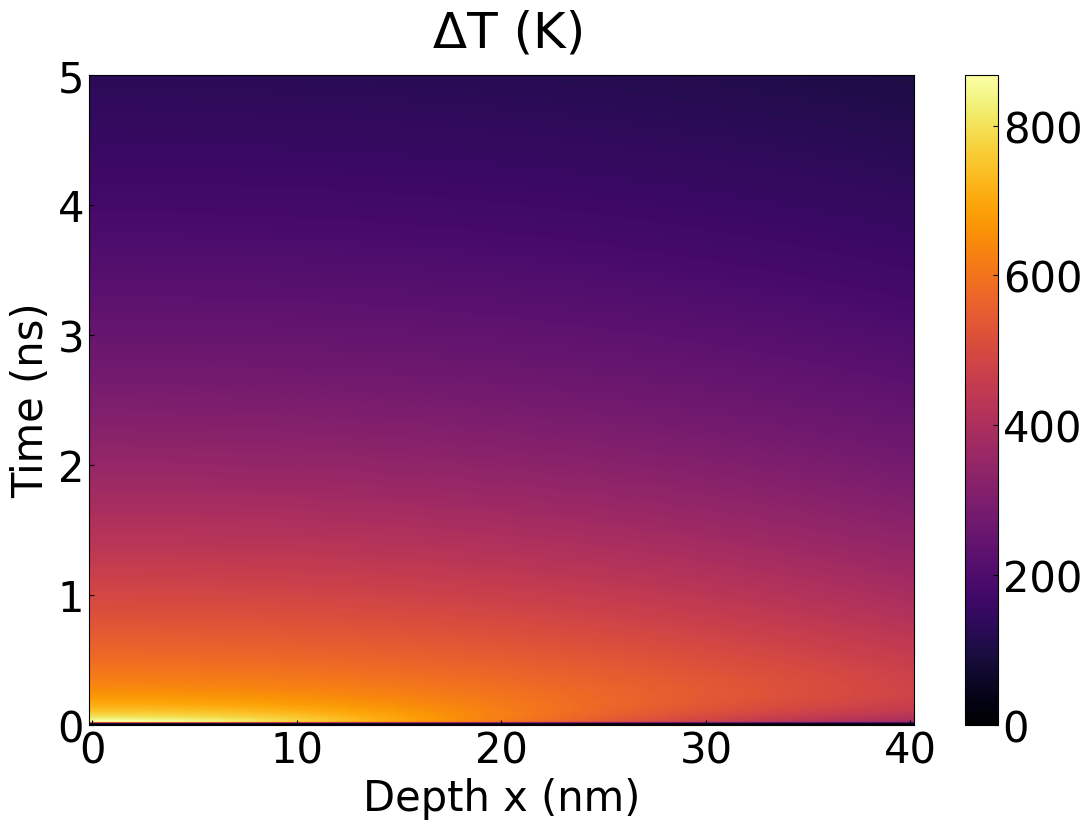}
  \caption{Calculation results of the one-dimensional heat-transport model for a Te thin film on sapphire.
  }
  \label{fig:pperp}
\end{figure}

The following parameters are used in the calculations: thermal conductivity $\kappa = 1.6~\mathrm{W\,m^{-1}\,K^{-1}}$ \cite{huang20202dmater},
specific heat $C = 0.202~\mathrm{J\,g^{-1}\,K^{-1}}$ \cite{thomson2001book},
and mass density $\rho = 6240~\mathrm{kg\,m^{-3}}$ \cite{thomson2001book}.
We consider an optical pulse of fluence $U=4~\mathrm{mJ\,cm^{-2}}$ ,
pulse width $\sigma = 5\times10^{-12}~\mathrm{s}$, and pulse center time
$t_0 = 3\sigma$. The Te thickness is $d = 40~\mathrm{nm}$.
Figure~\ref{fig:pperp} shows the calculation results.
We find the characteristic cooling time, or $\Delta t$, of the Te film is of the order of nanoseconds.

Next, we determine the characteristic time scale over which rotational motion becomes randomized.
Even after reaching the melting point, the orientation of a Te grain is not expected to instantaneously randomize but instead remains correlated with its initial state for a finite time.
This memory-loss time scale can be estimated from the rotational relaxation time.
We then interpret its inverse as the characteristic frequency of reorientation attempts, i.e. $\nu_0$.

In Te, the orientation of the chainlike structure must be considered a stick, not an arrow (as in magnetization).
A stick looks the same when flipped by $\pi$, so such a flip does not create a different state.
For this reason, we use a second-order relaxation time, i.e., the second rotational relaxation time, which properly captures this symmetry.

Motivated by the Stokes--Einstein-Debye relation\cite{debye1929polar}, we model a Te grain
as an effective sphere with radius $R$.
Using the average Te chain length of the $3.91$ bonds\cite{akola2012prb} and the Te interatomic distance $0.2835~\mathrm{nm}$ \cite{adenis1989actacryst}, we estimate a characteristic grain size of
$\sim 1~\mathrm{nm}$, and thus take $R$ to be of the order of $1~\mathrm{nm}$.

The rotational diffusion coefficient $D_r$ and the $\ell$-th rotational relaxation time $\tau_\ell$
are given by
\begin{gather}
D_r=\frac{k_{\mathrm B}T}{8\pi\eta R^{3}},\qquad
\tau_\ell=\frac{1}{\ell(\ell+1)D_r}
=\frac{8\pi\eta R^{3}}{\ell(\ell+1)k_{\mathrm B}T}.
\end{gather}
Here, $\eta$ is the dynamic viscosity, $k_{\mathrm B}$ is the Boltzmann constant, and $T$ is the
temperature. Using the kinematic viscosity $\nu=\eta/\rho$ in $T=800~\mathrm{K}$,
$\nu = 2.4\times 10^{-7}~\mathrm{m^2\,s^{-1}}$ \cite{li2005thermophysical}, we obtain
$\eta=\rho\nu \approx 1.5\times 10^{-3}~\mathrm{Pa\,s}$ by taking $\rho=6240~\mathrm{kg\,m^{-3}}$ \cite{thomson2001book}.

For the axial reorientation of Te, we consider $\ell=2$ and estimate $\tau_2 \simeq 0.6~\mathrm{ns}$,
which suggests that it takes of the orders of nanoseconds to induce grain reorientation.
The attempt frequency is therefore of the order of $\nu_0 \sim \tau_2^{-1}\sim \mathrm{ns^{-1}}$, which is equivalent to the inverse of the characteristic cooling time $\Delta t$.
We thus obtain $\nu_0 \sim \tau_2^{-1} \sim \Delta t^{-1} \sim \mathrm{ns^{-1}}$.

\clearpage
\subsection{\label{sec:app:materials} Material parameters of various systems}
Tables~\ref{table1} and ~\ref{table2} show the material properties of Te, BP, WTe$_2$ and SnSe.
These values were used in the calculation of Fig~\ref{fig:Usat}.
$\rho$ were calculated from the lattice constants, the molar mass, and the number of formula units per unit cell.
Because 
$V$ was taken to represent the smallest structural unit in a monolayer, half of the volume of the unit-cell was used.
Although 
$z$ can easily vary depending on the thickness thickness of the film $d$, here we present the value of $d=40$nm.
\begin{table}[h]
\centering
\small
\renewcommand{\arraystretch}{1.2}
\setlength{\tabcolsep}{4pt}
\begin{tabular}{lcccccc}
\hline
 & \makecell{$C$\\(J g$^{-1}$ K$^{-1}$)}
 & \makecell{$\rho$\\(kg m$^{-3}$)}
 & $\bar{\eta}$
 & $\Delta\eta$
 & \makecell{$V$\\(m$^3$)}
 & \makecell{$T_{\mathrm{melt}}$\\(K)}\\
\hline
Te & 0.202 \cite{thomson2001book}
   & 6240 \cite{thomson2001book}
   & 4.62 + 1.20 i\cite{ciesielski2018matscisemiproc}
   & 0.137 + 1.22 i\cite{guo2022nanoscale}
   & $1.67\times10^{-28}$ \cite{akola2012prb,thomson2001book,ciaaw2024misc}
   & 723 \cite{thomson2001book}\\
BP & 0.695 \cite{Stephenson1969BPHeatCapacity}
   & 2690 \cite{Brown1965BPStructure}
   & 3.62+ 0.153 i\cite{Lee2022BPPhaseRetarder}
   & -0.0943 + 0.233 i\cite{Lee2022BPPhaseRetarder}
   & $7.59\times10^{-29}$ \cite{Brown1965BPStructure}
   & 880 \cite{muhammad2024melting}\\
WTe$_2$ &0.177 \cite{Callanan1992WTe2HeatCapacity}
   & 9440 \cite{Brown1966WTe2Structure}
   & 3.59 + 1.59 i\cite{Munkhbat2022TMDOpticalConstants}
   & -0.383 + 1.07 i\cite{Munkhbat2022TMDOpticalConstants}
   & $1.55\times10^{-28}$\cite{Brown1966WTe2Structure}
   & 1293 \cite{Hansen2020WTe2PeritecticMelting}\\
SnSe & 0.263\cite{Wiedemeier1981SnSeHeatCapacity}
   & 6200 \cite{Chattopadhyay1986SnSeStructure}
   & 3.07+0.556 i\cite{Guo2023SnSePolarizer}
   & 0.114 + 0.165 i\cite{Guo2023SnSePolarizer}
   & $1.06 \times 10^{-28}$\cite{Chattopadhyay1986SnSeStructure}
   & 1146 \cite{feutelais1996phase}\\
\hline
\end{tabular}
\caption{Material parameters of Te, BP, WTe$_2$ and SnSe used in the calculations.}
  \label{table1}
\end{table}

\begin{table}[h]
\begin{tabular}{lccc}
\hline
 &$x$
 &$y$
 &\makecell{$z$ \\(cm$^2$ mJ$^{-1}$)}\\
\hline
Te & 21.7
   & 0.5051
   & 0.1563\\
BP & 20.06
   & 0.7614
   & 0.02803\\
WTe$_2$ &62.59
   &0.3365
   & 0.1675\\
SnSe & 35.60
   & 0.1484
   & 0.1058\\
\hline
\end{tabular}
\caption{Parameters $x,y,z$ calculated for Te,BP,WTe$_2$ and SnSe.}
  \label{table2}
\end{table}

\clearpage
\bibliography{refs_050826}

\end{document}